\documentclass[aps,prb,twocolumn,superscriptaddress]{revtex4-1}

\usepackage{graphicx}
\usepackage{amssymb}
\usepackage{amsmath}
\usepackage{amsfonts}
\usepackage{braket}
\usepackage{color}

\begin{document}

\title{Goldstone mode of Skyrmion Crystal  } 
\author{V.~E. Timofeev}
\email{Victor.Timofeev@thd.pnpi.spb.ru}
\affiliation{NRC ``Kurchatov Institute", Petersburg Nuclear Physics Institute, Gatchina
188300, Russia}
\affiliation{St.Petersburg State University, 7/9 Universitetskaya nab., 199034
St.~Petersburg, Russia} 
\author{D.~N. Aristov}
\affiliation{NRC ``Kurchatov Institute", Petersburg Nuclear Physics Institute, Gatchina
188300, Russia}
\affiliation{St.Petersburg State University, 7/9 Universitetskaya nab., 199034
St.~Petersburg, Russia} 
 
\begin{abstract}
We discuss the Goldstone mode of skyrmion crystal in a model of two-dimenssional ferromagnet with Dzyaloshinskii-Moriya interaction in  magnetic field. We use stereographic projection approach to construct skyrmion crystal and consider skyrmion's displacement field.  The small overlap of the individual skyrmion images restricts the potential energy to the interaction of nearest neighboring displacements.  The closed form of the Goldstone mode dispersion is found and its dependence on the magnetic field is studied. We use semiclassical quantization to define the Green's function and show that the propagation of displacements through the crystal changes its tensorial form from isotropic to anisotropic one at large distances. 
\end{abstract}

\maketitle

{\bf Introduction}. 
Magnetic skyrmions are topologically nontrivial whirls of local magnetization. They may serve as building blocks for novel racetrack memory devices\cite{Vakili_2021} or programmable logic devices\cite{Yan2021}, thanks to topological protection, small size and ability to be manipulated by spin torques.

One can consider a magnetic skyrmion\cite{Kiselev_2011} as an extremely small magnetic bubble\cite{Leeuw_1980} (cylindrical domain wall\cite{Thiele_1969}). In  terms of domain walls, skyrmion's radius is comparable with its width and defined by Dzyaloshinskii-Moriya interaction (DMI) constant\cite{Kiselev_2011,BOGDANOV1994255}. %Small size of skyrmions leads to interesting constraint for topological properties of SkX cell\cite{Bogatyrev2018}.

The dynamics of a local magnetization is usually  described by the Landau-Lifshitz-Gilbert(LLG) equation.  LLG equation can be further transformed into the Thiele\cite{thiele1973} equation in case of domain wall steady motion.  Thiele equation and its generalizations\cite{Weisenhofer2021} are the main tool for skyrmions motion analysis\cite{fert2017magnetic}. Thiele equation allows one to take into account the  spin current impact on  skyrmions and discuss the situation of skyrmions on a track, see \cite{fert2017magnetic}

In non-centrosymmetric magnets with DMI magnetic skyrmions are often arranged into regular lattices\cite{muhlbauer2009skyrmion,Yu_2010,Yu2010b}.
Such lattices (called also skyrmion crytals (SkX)\cite{Nagaosa2013}) are preferable to uniform, helix or cone configurations in case when a single skyrmion configuration becomes energetically more favorable \cite{timofeev2021}.   It was shown that the densely packed skyrmion configurations is characterized by both pairwise repulsive  and triple attractive interaction between
skyrmions\cite{Timofeev2019}. Hence, the motion of individual skyrmions in a lattice depends on its neighbours, and the SkX dynamics cannot generally be reduced to the motion of solitary skyrmion in a potential well.

Excitations of SkX have are described by the complicated band structure\cite{roldan2016,garst2017collective,PhysRevResearch.2.033491,Timofeev_2022}. The lattice excitations of different angular symmetry correspond to different distortions of individual skyrmions, among them elliptical deformation, breathing mode, clockwise and counter-clockwise motion etc. The modes with certain symmetries\cite{Timofeev_2023} show up in magnetic resonance experiments\cite{onose2012observation}.

%It is obvious that energy of SkX doesn't change after displacement of all skyrmions in SkX on arbitrary vector, thus there should be Goldstone mode. 
The soft Goldstone mode of SkX, also called gyrotropic mode, is associated with displacement of skyrmions in SkX and was predicted in Ref.\  \cite{petrova2011}. SkX was represented there as a sum of three magnetic helices with the corresponding  phase shifts, and it was shown that the topological term in the Lagrangian leads to quadratic dispersion of the soft mode (it was recently verified numerically in Ref.\ \cite{Mohanta2020}). The gyrotropic mode does not manifest itself in magnetic resonance experiments\cite{Schwarze2015}, but somehow appears in inelastic neutron scattering\cite{Weber_2022}.

%Existence of a gapless Goldstone mode with quadratic dispersion was predicated in the work\cite{petrova2011}, with assumption that SkX can be presented as a sum of three magnetic helices, 

In this work we consider a simplest model of non-centrosymmetric ferromagnet with DMI in external magnetic field, whose ground state is SkX  in a certain range of parameters. Staying in a framework of stereographic projection approach and regarding skyrmions in \emph{quasiparticles} paradigm, we consider a displacement field of skyrmions positions in the lattice. We numerically show the nearest-neighbour character of displacements' coupling and obtain closed form  of dispersion of the Goldstone mode. The dependence of {\it force constants} on external magnetic field is also numerically found. The dynamical Green's function of displacements is isotripic at small distances, while showing anisotropic tensor structure at larger distances. 

{\bf Model}. We consider planar model of non-centrosymmetric ferromagnet with DMI in uniform external magnetic field perpendicular to the plane. The energy density is given by:
\begin{equation}
\mathcal{E} =   \frac{C}{2}  \partial_{\mu}S_{i}\partial_{\mu}S_{i} - 
D\epsilon_{\mu ij} S_{i}\partial_{\mu}S_{j}  - B  S_{3},
\label{classicalenergy}
\end{equation}
where $C$ is an exchange parameter, $D$ is DMI constant, and $B$ is an external magnetic field magnitude. There is a convenient way to choose measurement units in the model \eqref{classicalenergy}: we will measure length in the units of $l=C/D$, and energy density in the units of $ CS^2 l^{-2} = S^2D^2/C$. Then the energy of the model depends only on the dimensionless parameter $b=BC /SD^2$. We consider low temperature limit, when local magnetization is saturated, and its magnitude doesn't change from point to point $\mathbf{S} = S \mathbf{n}$, with $|\mathbf{n}|=1$. The above planar model is applicable also to thin films, whose thickness is less or comparable to $l$. We ignore the magnetic dipolar interaction here, because it can be reduced to uniaxial anisotropy for  ultrathin films. Small anisotropy leads only to minor changes of SkX parameters.

The stereographic projection representation of the vector $\mathbf{n}$ reads as 
\begin{equation}
n_1 + i n_2  = \frac{2f}{1 + f\bar{f}}\,,\quad 
n_3 = \frac{1 - f\bar{f}}{1 + f\bar{f}},
\label{eq:stereo}
\end{equation}
with $f$ a complex-valued function, and $\bar{f}$ its complex conjugate. A single skyrmion's stereographic function is conveniently represented by 
\begin{equation}
f_1=\frac{i \, z_0 \,\kappa(z\bar{z}/z_0^2)}{\bar{z}},
\label{eq:anz}
\end{equation}
where $\kappa$ is a smooth real profile function, $z_0$ is a skyrmion size parameter. The ansatz \eqref{eq:anz} is more convenient for the  description of SkX case, while in case of one skyrmion it reproduces the profile obtained by usual bubble domain ansatz.  It was shown previously that the multi-skyrmion configurations can be built as a sum of stereographic functions of individual skyrmions \cite{Timofeev2019}. Particularly, the regularly arranged  SkX corresponds to the following stereographic function:
\begin{equation}
f_{SkX} = \sum\limits_{n,m} f_1 (\mathbf{r} - n \mathbf{a}_1 - m \mathbf{a}_2),
\label{SkXf0}
\end{equation}
where $\mathbf{a}_1 = (0,a)$, $\mathbf{a}_2 = (-\sqrt{3}a/2,a/2)$, and $a$ is a cell parameter of SkX. The static properties of this ansatz \eqref{eq:anz}-\eqref{SkXf0} was discussed to some detail in previous works\cite{Timofeev2019,timofeev2021}. It was shown that the proposed SkX configuration has lower energy than helix or uniform configuration at magnetic fields, $0.25 \alt b \alt 0.8$.

\begin{figure}[t]
\center{\includegraphics[width=0.99\linewidth]{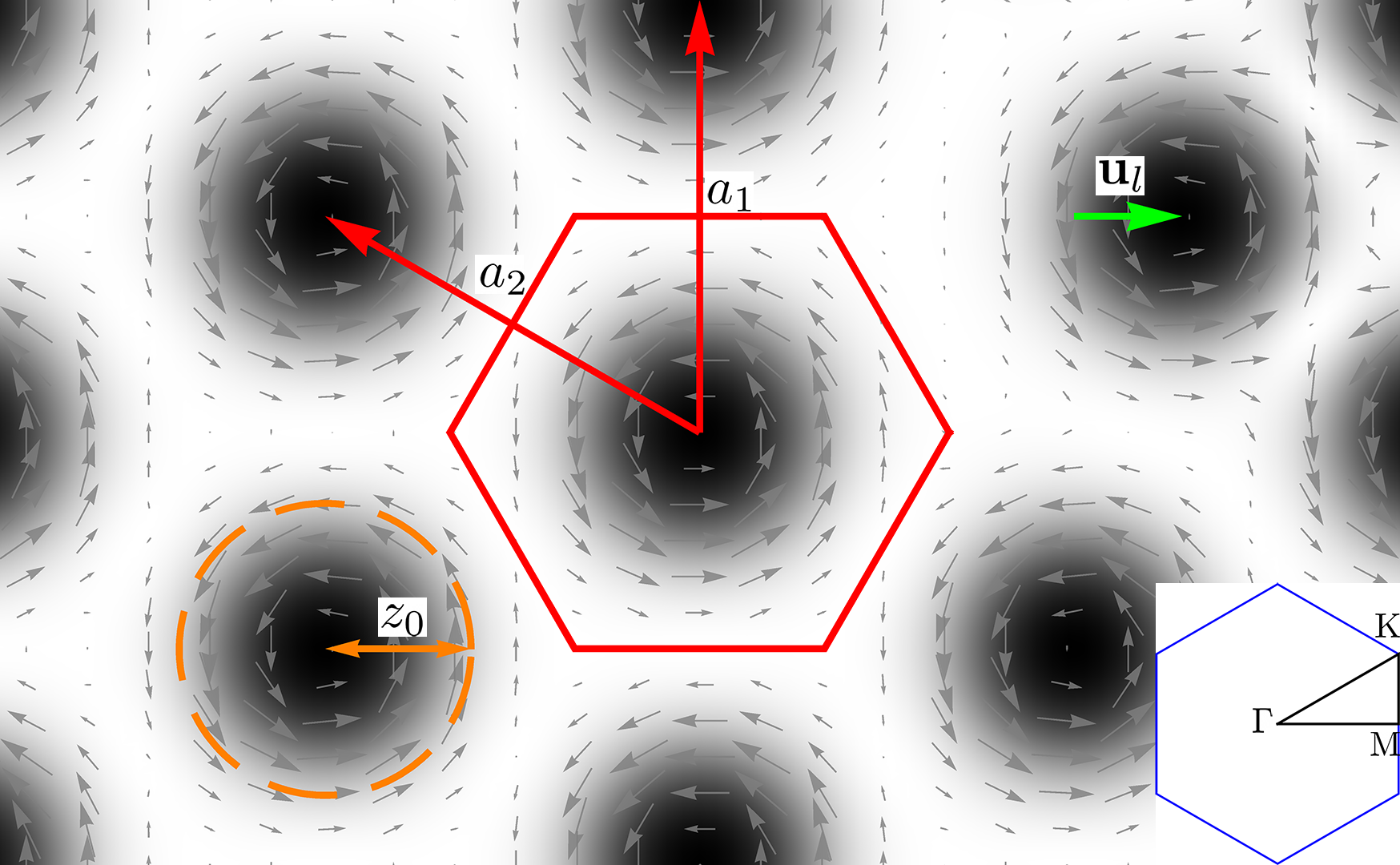}}
\caption{A sketch of SkX with one displaced skyrmion. Red arrows and a hexagon illustrate lattice vectors and the primitive cell of SkX, the orange circle shows a typical value of $z_0$ parameter, green arrow indicates a displacement of skyrmion in the right top corner. The Brillouin zone with symmetry points is depicted in the bottom right corner.}
\label{fig:sus}
\end{figure}

The dynamics of local magnetization follows from the Lagrangian, $ \mathcal{L} = \mathcal{T} - \mathcal{E}$,  with the  kinetic term 
\begin{equation}
\mathcal{T}=% \tfrac 12  
\frac { S}{ \gamma_0}(1-\cos{\theta})\dot{\varphi}\,,
\label{eq:kinetic}
\end{equation}
here  $\varphi$ and $\theta$ define the  magnetization direction $\mathbf{n}=(\cos\varphi\sin\theta,\sin\varphi\sin\theta,\cos\theta)$, and $\gamma_0$ gyromagnetic ratio. This form of the Lagrangian leads to the well known Landau-Lifshitz equation.  The expression \eqref{eq:kinetic} may be rewritten in terms of  $f$ as 
\begin{equation}
\mathcal{T}[f]=  \frac i2 \frac{\bar f \partial_t f - f \partial _t \bar f}
{1+f \bar f}\,, 
\label{kinLagrangian}
\end{equation}
and the factor $S/ \gamma_0$ was included into the time scale. The exact equation of motion for $f(t)$ is highly nonlinear and cannot generally be solved. In previous works \cite{Timofeev_2022,Timofeev_2023} we have discussed the normal modes of infinitesimal fluctuations of the function $f$. In this work we develop a special approach for consideration of the gyrotropic mode of SkX.\\

%%%%%%%%%%%%%%%%%%%%%%%%%%%

{\bf Displacement field and dispersion}. In our previous papers we associated the above configuration $f_{SkX}$, Eq.\eqref{SkXf0}, with the equilibrium spin configuration, $f_0$,  and  
considered small fluctuations, $\delta f$, around it, writing 
\[ 
f = f_0 + \delta f \equiv f_0 + (1+f_0 \bar{f}_0) \psi \, , \]
with the dynamics of $\psi$ is discussed at length in \cite{Timofeev_2022}. 

In this study we assume that the skyrmion lattice is imperfect, in the sense that the stereographic image is still given by the sum of individual images of skyrmions, and the shape of each image is unchanged, but the only imperfection is the position of the center of skyrmions.
It is similar to the description of ions' displacements in crystals  in the theory of phonons. 

\begin{equation}
f_{SkX}  = \sum\limits_{l} f_1 (\mathbf{r} - \mathbf{r}^{(0)}_l  +  \mathbf{u}_l),
\label{SkXf}
\end{equation}
where $\mathbf{r}^{(0)}_l = n \mathbf{a}_1 + m \mathbf{a}_2$ with integer $n, m$ and  $\mathbf{a}_{1,2}$ lattice vectors. For infinitesimal displacements $\mathbf{u}_l$ we can write 

\begin{equation}
    f_{SkX}  \simeq f_{0} +  \sum\limits_{l} \mathbf{u}_l \nabla f_1 (\mathbf{r} - \mathbf{r}^{(0)}_l),
%\label{SkXf}
\end{equation}
In order to make use of our previously found formulas  in \cite{Timofeev_2022}, we define the quantity $\psi(\mathbf{r} ) $ as  
% Eq(13)

\begin{equation}
    \sum\limits_{l} \mathbf{u}_l \nabla f_1 (\mathbf{r} - \mathbf{r}^{(0)}_l)
  =   (1+f_0 \bar f_0)\, \psi(\mathbf{r} )   ,
%\label{SkXf}
\end{equation}
also 
$\mathbf{u}_j \nabla = u^+_j \partial_{z}  + u^-_j \partial_{\bar{z}}$ with 
 $\partial_{z} = (\partial_{x}-i\partial_{y})/2$ , $\partial_{\bar{z}} = (\partial_{x}+i\partial_{y})/2$ and $u^\pm_j = u^x_j \pm i u^y_j$. 
Introducing shorthand notation $f_j =  f_1 (\mathbf{r} - \mathbf{r}^{(0)}_j)$  we can write 

\begin{equation}
\psi =   \sum\limits_{j} 
\frac { u^+_j \partial_{z} f_j  + u^-_j \partial_{\bar{z}} f_j   }{1+f_0 \bar f_0}  , 
%\label{SkXf}
\end{equation}
and similarly for complex conjugated $\bar \psi$. 
As a result, we obtain 
\begin{equation}
\begin{aligned}
         \begin{pmatrix}
           \psi \\ \bar \psi
       \end{pmatrix} 
       &= \frac1{1+f_0 \bar f_0} \sum_j \begin{pmatrix}
      \partial_{\bar z} f_j ,& \partial_{ z} f_j \\ \partial_{\bar z} \bar f_j  ,& \partial_{ z} \bar f_j 
    \end{pmatrix}  
    \begin{pmatrix}
         u^-_j \\ u^+_j
       \end{pmatrix} 
    \,, \\ 
    & \equiv  \sum_j {\cal O}_j \begin{pmatrix}
         u^-_j \\ u^+_j
       \end{pmatrix}  \,.
\end{aligned}
\label{zeromode}
\end{equation} 
For the uniform shift $\mathbf{u}_j = \mathbf{u} $, using the property  $\sum _j ( f_j , \bar f_j ) = (f_0, \bar f_0)$, we restore the previously discussed zero modes (Eqs. (29) in \cite{Timofeev_2022} ) : 

\begin{equation}
\begin{aligned}
       \Psi_{\mbox\o}  &= \frac1{1+f_0 \bar f_0}\begin{pmatrix}
      \partial_{\bar z} f_0 \\ \partial_{\bar z} \bar f_0
    \end{pmatrix} \,, \\
    \bar\Psi_{\mbox\o}  &= \sigma_1  \Psi_{\mbox\o} ^* = 
    \frac1{1+f_0 \bar f_0}\begin{pmatrix}
      \partial_{z} f_0 \\ \partial_{z} \bar f_0
    \end{pmatrix} \,.
\end{aligned}
% \label{zeromode}
\end{equation}

The quadratic in displacements part of the  Lagrangian takes the form  

\begin{equation}
\begin{aligned}
\mathcal{L} & =\frac12 
 \sum_{lj} \begin{pmatrix}
         u^+_l ,&  u^-_l
       \end{pmatrix} \left( -i \hat{\mathcal K}_{lj}  \partial_t  
       -\hat{\mathcal{H}}_{lj} \right) 
       \begin{pmatrix}
         u^-_j \\   u^+_j
       \end{pmatrix} 
      \, ,  \\ 
      \hat{\mathcal K}_{lj} & = 
      \int d\mathbf{r}\, {\cal O}^\dagger_l  . \sigma_3 .  {\cal O}_j 
      \, ,  \\ 
        \hat{\mathcal H}_{lj} & =  
     \int d \mathbf{r}\, {\cal O}^\dagger_l .   \begin{pmatrix}
  (-i\nabla + \mathbf{A})^2 + U&
   V\\
  V^*&
   (i\nabla + \mathbf{A})^2 + U
\end{pmatrix} .{\cal O}_j  \,, 
\end{aligned} 
\label{eq:Lagr2}
\end{equation}
 with the explicit form of  $U$, $V$ and $\mathbf{A} $ is given in  \cite{Timofeev_2022}. 
 \footnote{
 We take the opportunity to correct the misprint in \cite{Timofeev_2022}, the definition of $\mathbf{A} $ there should contain overall minus sign and the factor 2 instead of 4 in the term in curly brackets. }

Let us discuss a few general properties. 

(i) Due to translation invariance, the quantities $\hat{\mathcal K}_{lj}$ and $  \hat{\mathcal H}_{lj}$ depend only on the difference $\mathbf{d}=\mathbf{r}^{(0)}_l -\mathbf{r}^{(0)}_j$. 

(ii) The zero mode corresponds to summation over $j$, and we should have  $ \sum_j \hat{\mathcal H}_{lj} = \sum_l   \hat{\mathcal H}_{lj} = 0 $, see  below.  

(iii) The image of a single skyrmion $f_1 (\mathbf{r})$ decreases exponentially  with distance. It follows that $\hat{\mathcal K}_{lj}$,  $  \hat{\mathcal H}_{lj}$ decrease rapidly with $|\mathbf{r}^{(0)}_l -\mathbf{r}^{(0)}_j|$. For practical reasons it suffices to consider only on-site term, $l=j$, and the nearest neighbors (NN).   

(iv)
It can be shown that  $ \sum _{j}\hat{\mathcal K}_{lj} = \pi \sigma_3 $, it corresponds to the value of topological charge per unit cell of skyrmion crystal. 

(v)
For triangular lattice with six NN we have $\mathbf{d}=  (a\,\cos \phi_d, a\,\sin \phi_d) $ with $\phi_d = \frac \pi 3 (n-1/2)$ and $n=0,\ldots 5$. Individual skyrmions are characterized by certain chirality, $f_1(\mathbf{r}) \propto e^{i\phi} /r $. Considering the symmetry of the potentials $U,V$ and matrix $ {\cal O}_j$ under the rotation $\phi \to \phi+\phi_d$, we notice that  the phase $\phi_d$ doubles in the off-diagonal components  and is absent in diagonal components of $\hat{\mathcal H}_{lj}$. As a result we  have a structure 
\[ 
\hat{\mathcal H}_{lj} =  \begin{pmatrix}
    h_1, & h_2 e^{-2i\phi_d} \\ h_2 e^{2i\phi_d}, & h_1 
\end{pmatrix} \,, 
\]
with $h_{1,2}$ depending only on the distance, $d$. 
For the on-site term $l=j$, the off-diagonal components are absent, $h_2=0$. Numerically, we find that $h_{1,2} < 0$ for $l \neq j$.   

\begin{figure}[t]
\center{\includegraphics[width=0.99\linewidth]{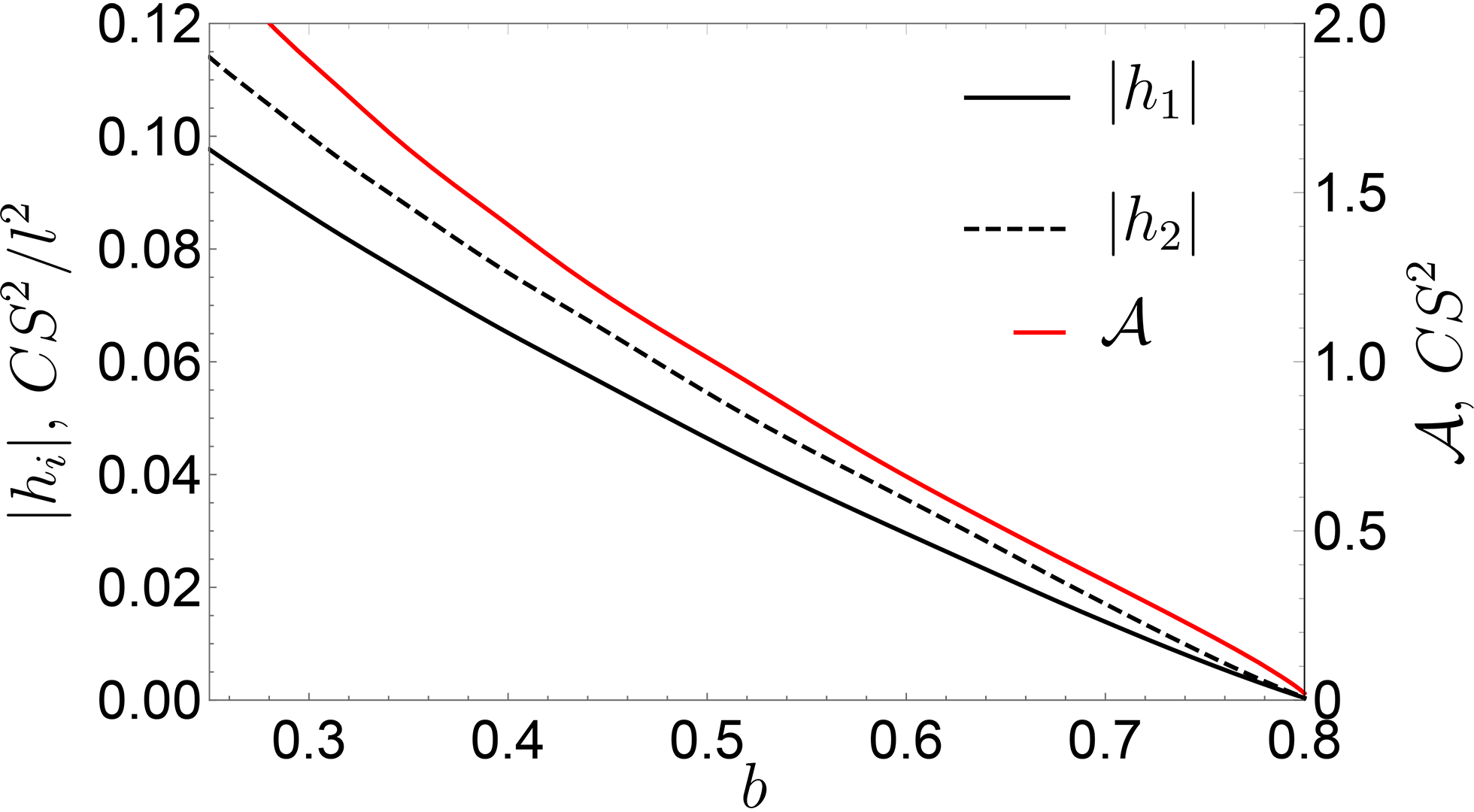}}
\caption{Dependence of the hopping constants, $h_1$, $h_2$, and the stiffness, ${\cal A}$, on magnetic field $b$.}
\label{fig:hop}
\end{figure}

(vi) The Thiele equation for the motion of $l$'th skyrmion is obtained by putting $u^\pm_j = 0$ in \eqref{eq:Lagr2} for all $j\neq l$. In this case the Lagrangian becomes ${\cal K}_{ll} u_l^x {\dot u}_l^y - h_1 ((u_l^x)^2 + (u_l^y)^2)$, cf.\ \cite{metlov13vortex}.  
Notice that even if $u^\pm_j = 0$ initially, the collective character of Eq.\ \eqref{eq:Lagr2} leads to eventual propagation of perturbation around the initial displacement. We discuss it in more detail below.

Using the above  properties, 
we represent the quadratic part of the Lagrangian as 
 \begin{equation}     \begin{aligned}
   \mathcal{L}   &=    \frac12  \sum _\mathbf{q} 
  \begin{pmatrix}          u^+_{-\mathbf{q}} ,&  u^-_{-\mathbf{q}}        \end{pmatrix} 
  \left( -i \hat{\mathcal K}_{\mathbf{q}}  \partial_t         -\hat{\mathcal{H}}_{\mathbf{q}} \right) 
       \begin{pmatrix}         u^-_\mathbf{q} \\   u^+_\mathbf{q}        \end{pmatrix} 
      \, ,  \\  
      \hat{\mathcal K}_{\mathbf{q}}  &= (\pi + k_1\gamma_s(\mathbf{q})  )\sigma_3  \,, \\ 
     \hat{\mathcal H}_{\mathbf{q}} &= \begin{pmatrix}
    h_1\gamma_s(\mathbf{q}) , & h_2\,  \gamma_d^*(\mathbf{q})  \\ h_2\, \gamma_d(\mathbf{q}), & h_1\gamma_s(\mathbf{q})  
\end{pmatrix} \,,    
    \end{aligned}   
    \label{eq:L2}
\end{equation}
where we defined  $ u^x_j \pm i u^y_j = \sum _\mathbf{q} e^{i \mathbf{q} \mathbf{r}_j} u^\pm_\mathbf{q}  \,, $  
and  the sums over six NN are  
  \begin{equation}     \begin{aligned}
    \gamma_s(\mathbf{q})  &=  \sum  _\mathbf{d} e^{-i \mathbf{q} \mathbf{d}} - 6  
    \\ & = 
    2\left( 2  \cos \tfrac{ \sqrt{3}} 2   q_x a \cos \tfrac12  {q_y a}   +  \cos q_y a  -3  \right)
     \, , \\ 
    % \gamma_p(\mathbf{q})  
    % &=   \sum _\mathbf{d} e^{-i \mathbf{q} \mathbf{d}} e^{ i\phi_d} 
    % \\  & = 
    % -2 \left(   \cos \tfrac{ \sqrt{3}} 2   q_x \sin \tfrac12  {q_y}   +  \sin q_y 
    % \right. \\ &  \left.
    % -   i   \sqrt{3}  \sin \tfrac{ \sqrt{3}} 2   q_x \cos \tfrac12  {q_y} \right) 
    % \, , \\ 
    \gamma_d(\mathbf{q}) % t_2^- (\mathbf{q})
    &=   \sum _\mathbf{d} e^{-i \mathbf{q} \mathbf{d}} e^{ 2i\phi_d} 
    \\  & = 
    2 \left(   \cos \tfrac{ \sqrt{3}} 2   q_x a \cos \tfrac12  {q_y a}   -  \cos q_y a 
    \right. \\ &  \left.
    -   i   \sqrt{3}  \sin \tfrac{ \sqrt{3}} 2   q_x a \sin \tfrac12  {q_y a} \right)
     \,,   
    \end{aligned}   
    \label{defgamma}
\end{equation}
with the property  $\gamma_{s,d}(0)  = 0$.  
% Here we put $d=1$ for simplicity of notation. 
The zero value of diagonal components  of $\hat{\mathcal H}_{\mathbf{q}} $ at $\mathbf{q}=0$ 
is the explicit use of  the above property (ii), and  we will return to it below. 

The dispersion law $\omega = \epsilon_{\mathbf{q}}$ in the time dependence $  u^\pm_\mathbf{q} (t) = e^{i\omega t}  u^\pm_\mathbf{q}  $ is given by solving the equation 
$\mbox{det} (\omega \hat{\mathcal K}_{\mathbf{q}}  -\hat{\mathcal{H}}_{\mathbf{q}} ) =0$, which yields
\[ \epsilon_{\mathbf{q}}  = \frac { (h_1^2 \gamma_s^2(\mathbf{q}) - h_2^2  |\gamma_d(\mathbf{q})|^2)^{1/2} }{\pi + k_1 \gamma_s(\mathbf{q}) } \]
% \[ \epsilon_{\mathbf{q}}  = \frac { (h_1^2 \gamma_s^2  - h_2^2  |\gamma_d |^2)^{1/2} }{\pi + k_1 \gamma_s  } \]

\begin{figure}[t]
\center{\includegraphics[width=0.99\linewidth]{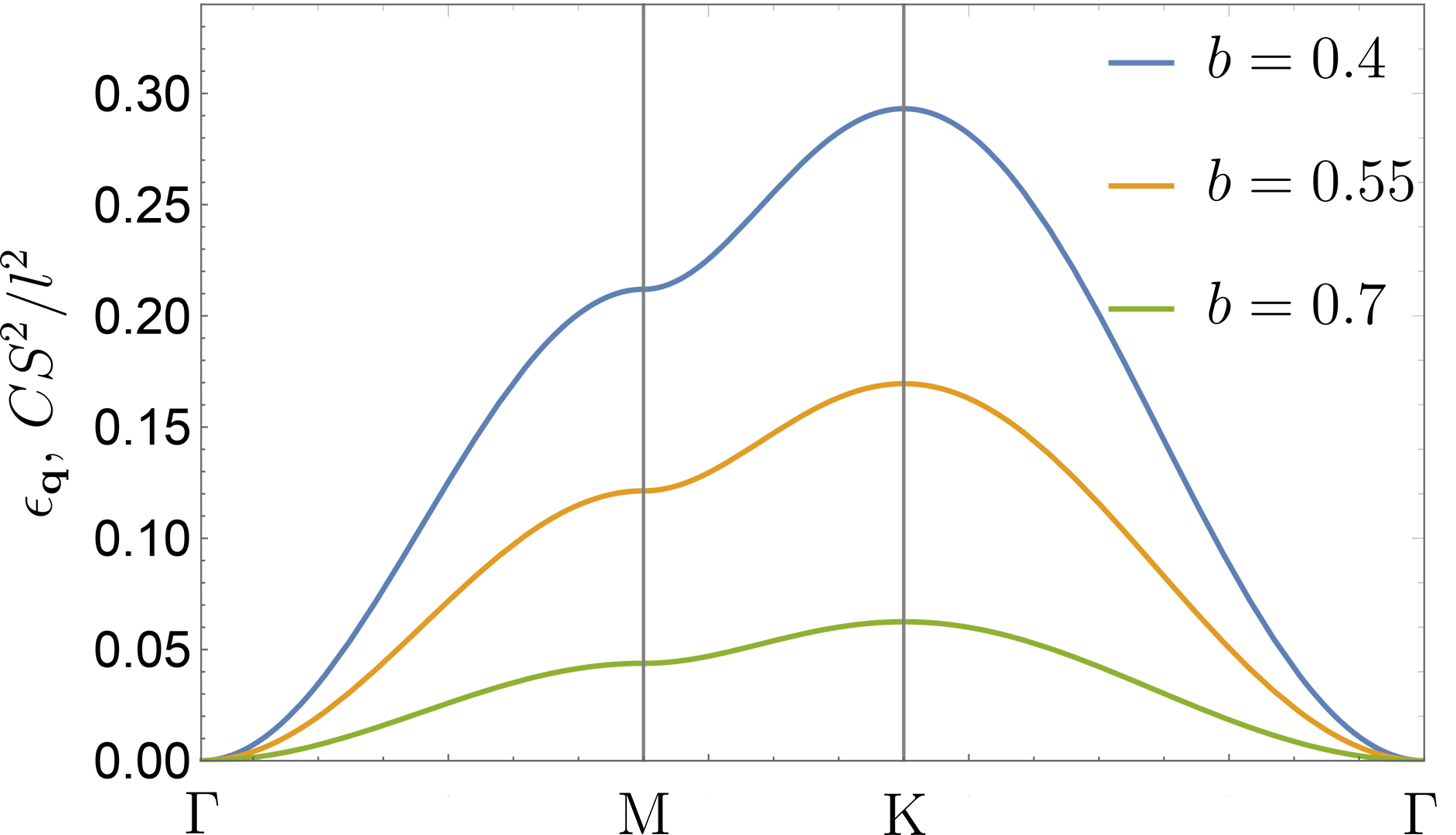}}
\caption{Dispersion $\epsilon_{\mathbf{q}}$ for  different $b$ values. Symbols $\Gamma$, M, K correspond to symmetry points in Brillouin zone, depicted in Fig.\,\ref{fig:sus}.}
\label{fig:disp}
\end{figure}

We obtain numerically that $h_1\simeq 0.84\,  h_2$ in the whole range of relevant fields, $b\in (0.3, 0.8)$.
The dependence of $h_{1,2}$ on $b$ is shown in Fig.\  \ref{fig:hop}.
We see here that both coefficients  vanish simultaneously at the critical value of the field, $b_c \simeq 0.8$. 
In terms of elastic theory, discussed in  \cite{petrova2011}, it  corresponds to Lam\'e coefficients $\lambda$, $\mu$ decreasing and vanishing at $b=b_c$, whereas $\lambda \simeq 1.94\mu$, see also Eq.\ \eqref{eq:L3} below. 
The dependence of $\epsilon_{\mathbf{q}}$ along the symmetry lines of the Brillouin zone is shown in Fig.\ \ref{fig:disp}. 

The values of $\epsilon_{\mathbf{q}}$ are $9 |h_1|/\pi$ and $2\sqrt{4h_1^2- h_2^2}/\pi$ at $\mathbf{q} = K$ and $\mathbf{q} = M$, respectively, 
and we have $\epsilon_{\mathbf{q}=M} / \epsilon_{\mathbf{q}=K} \simeq 0.71$ for all values of $b$. 
The diminishing of the gyrotropic bandwidth with $b$ while maintaining the above ratio $0.71$ is consistent with the results reported in \cite{PhysRevResearch.2.033491}.
The amplitude $k_1$ for the hopping contribution in the kinetic term $ \hat{\mathcal K}_{\mathbf{q}} $ is negative and small, $|k_1| < 2\cdot 10^{-2}$, and we can safely ignore it in a qualitative discussion below. 

 In the  limit of  small wave vectors, $|q|\ll 1$, we have 
\begin{equation}
\epsilon_{\mathbf{q}} \simeq   \tfrac3{4\pi} q^2 a^2\sqrt{4h_1^2 - h_2^2} \equiv {\cal A} q^2 \,,
\label{Esmallq}
\end{equation}
The dependence of the stiffness coefficient, ${\cal A}$, on $b$ is also shown in Fig.\  \ref{fig:hop} ; there is no simple proportionality between  $h_{1,2}$ and ${\cal A}$, since $d$ also depends on $b$, see \cite{Timofeev_2022}. 
It is worthwhile to compare  the value of ${\cal A}$ with the  dispersion law of uniform ferromagnet. 
The latter case corresponds to $f_0\equiv 0$, and further to $U=b$, $V=0$ and $\mathbf{A} =0 $ in \eqref{eq:Lagr2}, see Eqs.\  (16)-(18) in \cite{Timofeev_2022}. This results to  $\epsilon_{\mathbf{q}} = q^2 + b $, i.e.\ to ${\cal A}=1$ and the gapped character of the spectrum. We see that compared to the uniform ferromagnetic case, the lowest-lying part of the spectrum of SkX is gapless and its stiffness decreases with the field. One may ask, how this gapless character becomes gapful one at $b_c$, when the SkX dissolves. The answer is that the applicability of \eqref{Esmallq}  shrinks to $q=0$  in this case, as $a\to \infty$ at $b_c$, cf.\  \cite{Aristov02}.

%%%%%%%%%%%%%%
%  justification of zero 
%%%%%%%%%%%%%%
Let us now discuss  the above property (ii), which we incorporated into the definition of $\gamma_s(\mathbf{q})$.  Our numerical computation with the use of our trial function shows that the property $\hat{\mathcal H}_{\mathbf{q}} =0 $ at $\mathbf{q}=0$ is satisfied to the accuracy of $5\cdot 10^{-3}$ in the range of the fields, $0.35<b<0.75$. This shows rather good quality of our trial function, in accordance with our previous paper \cite{Timofeev_2022}. At the same time, since our trial function does not provide the exact extremum of the action,  its first variation is not  identically zero. In this case additional terms should be added to  the effective Hamiltonian, $\hat{\mathcal H}_{lj} $ in  \eqref{eq:Lagr2}, as is explained below. 

We expand the variation of our function to second order in displacements, 
 \begin{equation}
\begin{aligned}
  \delta f & = (1+f_0\bar f_0)^{-1} \psi = 
\sum\limits_{l} \left(  {u}^\alpha_l \partial_\alpha f_1 (\mathbf{r} - \mathbf{r}^{(0)}_l)  
\right.   \\   & \left. 
 + \tfrac12 {u}^\alpha_l  {u}^\beta_l   \partial_\alpha \partial_\beta f_1 (\mathbf{r} - \mathbf{r}^{(0)}_l)
\right) 
\end{aligned}
\label{SecondOrder}
\end{equation} 
This expression should be multiplied by $\delta {\cal L}/\delta f$, which is now assumed not being identically zero.
The first order term here, $\propto {u}^\alpha_l$, when integrated over $\mathbf{r}$, produces the force applied to $l$th skyrmion out of equilibrium position. 
The second-order term in \eqref{SecondOrder}, when multiplied by $\delta {\cal L}/\delta f$ and integrated over $\mathbf{r}$, adds to expression \eqref{eq:Lagr2} in a following way. 

First we note that the kinetic part of $\delta {\cal L}/\delta f$, being symmetric in indices $\alpha$, $\beta$, leads to terms proportional to full time derivatives, $\propto   \tfrac{d}{dt}({u}^\alpha_l)^2$, which can be discarded. The potential part of the Lagrangian can be integrated by parts and the boundary term vanishes due to the rapid decrease 
of $f_1 (\mathbf{r})$ with distance. The rest can be represented, after some calculation, in a form  
\begin{equation}
\begin{aligned}
\mathcal{L}' & = \frac12 
 \sum_{lj} \begin{pmatrix}
         u^+_l ,&  u^-_l
       \end{pmatrix}     
       \hat{\mathcal{H}}_{lj}   
       \begin{pmatrix}
         u^-_l \\   u^+_l
       \end{pmatrix} 
      \, , 
\end{aligned} 
\label{eq:Lagr2a}
\end{equation}
which means that we should replace $\hat{\mathcal{H}}_{lj}$ in  \eqref{eq:Lagr2} by $\tilde{\mathcal{H}}_{lj} =  \hat{\mathcal{H}}_{lj} - \delta_{lj} \sum_m \hat{\mathcal{H}}_{lm}$. Clearly,  if the property $\sum_j\hat{\mathcal{H}}_{lj}  = 0 $ is not fulfilled, due to imprecise character of our trial function, then this property is ultimately restored for the corrected form, $\tilde{\mathcal{H}}_{lj}$, with $\sum_j \tilde{\mathcal{H}}_{lj} =0$. Using the symmetry of   \eqref{SecondOrder} in indices $\alpha$, $\beta$, one can also show  that $\sum_l \tilde{\mathcal{H}}_{lj} =0 $ as well. It justifies the above general property (ii) and the subtraction of number $6$ in our definition of  $\gamma_s(\mathbf{q})$. \\

{\bf Green's function in small $q$ limit. } 
%%%%%%%%%%%%
% small q limit 
%%%%%%%%%%%%
Let us  discuss now the propagation of   displacements through the SkX. 
% The peculiarity of \eqref{Esmallq} is clarified when 
To simplify our discussion, we perform two subsequent rotations of our Lagrangian. 
First, we return to the Cartesian basis   $ u^\pm_\mathbf{q}  =   u^x_\mathbf{q} \pm i u^y_\mathbf{q}  $, which corresponds to rotation
$U = \begin{pmatrix} 1, & -i \\ 1, & i \end{pmatrix}$. 
Second, we 
bring $\hat{\mathcal H}_{\mathbf{q}}$ to principal axes, parallel and perpendicular to $\mathbf{q} = q (\cos \phi_q,\sin \phi_q)$, by writing  $(u^x_\mathbf{q}, u^y_\mathbf{q}) = (u^\|_\mathbf{q}, u^{\perp}_\mathbf{q})\cdot U_1^\dagger$ with    
with $U_1 =  \begin{pmatrix}   \cos \phi_q, & -\sin \phi_q \\ \sin \phi_q , &\cos \phi_q \end{pmatrix}$. 
As a result we reduce the Lagrangian to the form  
 \begin{equation}     \begin{aligned}
   \mathcal{L}   &=    \frac12  \sum _\mathbf{q} 
  \begin{pmatrix}    u^\|_\mathbf{-q}, & u^{\perp}_\mathbf{-q}     \end{pmatrix} 
   \begin{pmatrix} -A_{\|} q^2, & -2\pi \partial_t \\  2\pi \partial_t  ,&- A_{\perp}q^2 \end{pmatrix}
       \begin{pmatrix}         u^\|_\mathbf{q} \\   u^\perp_\mathbf{q}        \end{pmatrix} 
      \, ,  \\ 
      A_{\|} & = -\tfrac32  (2h_1  + h_2) a^2 \,, \quad 
     A_{\perp} = -\tfrac32  (2h_1  - h_2) a^2
     \,,         
    \end{aligned}   
    \label{eq:L3}
\end{equation}
with    $A_{\|} , A_{\perp} >0$  are two elastic moduli, that would correspond to longitudinal and transverse sound modes in situation with phonons; we have $A_{\|} A_{\perp} =2\pi {\cal A}$. The different form of the kinetic term in our case makes the  dispersion quadratic, instead of linear dispersion law for acoustic phonons; it also   shows that the displacement $2\pi\, u^\perp_\mathbf{-q}$ is a canonically conjugate momentum to  $ u^\|_\mathbf{q}$. 

It allows us to second quantize our theory, along the guidelines in \cite{Rajaraman,Timofeev_2022}. We demand $[u^\|_\mathbf{q},2\pi\, u^\perp_\mathbf{-q}]=i \hbar$ (we set $\hbar=1$) and find  
 \begin{equation}     \begin{aligned}
     u^\|_\mathbf{q} &= \frac{1}{\sqrt{4\pi \varkappa}}
     (c_\mathbf{q}^\dagger e^{i \epsilon_{\mathbf{q}}t} + c_\mathbf{-q}  e^{-i  \epsilon_{\mathbf{q}}t} )
     \, , \\ 
     u^\perp_\mathbf{q} &=i \frac{\sqrt{\varkappa}}{\sqrt{4\pi }}
    (c_\mathbf{q}^\dagger e^{i  \epsilon_{\mathbf{q}}t} - c_\mathbf{-q}  e^{-i  \epsilon_{\mathbf{q}}t} )
     \, , 
    \end{aligned}    
\end{equation}
where asymmetry parameter, $\varkappa =\sqrt{ A_{\|}/ A_{\perp} } \simeq 1.98$. In terms of creation (annihilation) operators, $c_\mathbf{q}^\dagger$ ($c_\mathbf{q}$), the Hamiltonian becomes $\hat{\mathcal H} = \sum_{\mathbf{q}} \epsilon_{\mathbf{q}} c_\mathbf{q}^\dagger c_\mathbf{q}$. 

% We perform further rotation in plane, $\hat{\mathcal H}_{\mathbf{q}} \to U_1^\dagger \hat{\mathcal H}_{\mathbf{q}} U_1$
% % with $U_1 =  \begin{pmatrix} \hat q_1, & -\hat q_2 \\ \hat q_2 , &\hat q_1 \end{pmatrix}$ and $\hat q_ j = q_j / q $. 
% with $U_1 =  \begin{pmatrix}   \cos \phi_q, & -\sin \phi_q \\ \sin \phi_q , &\cos \phi_q \end{pmatrix}$. 
% It  brings $\hat{\mathcal H}_{\mathbf{q}}$ to principal axes, parallel and perpendicular to $\mathbf{q}$, and leads to the equation 
% \[
% \mbox{det}   \begin{pmatrix}    A_{\|} q^2 , & -2 i \pi \omega  \\ 
% 2 i \pi \omega ,  &   A_{\perp} q^2 \end{pmatrix} 
%   = 0 \,,
% \]
% corresponding to above Eq.\ \eqref{Esmallq}.  

%Interestingly, we find that  $ A_{\|} < A_{\perp}$, in contrast to reversed relation expected for phonons.  

% The dependence of two elastic moduli,  $ A_{\|}$, $A_{\perp}$, on the value of the field, $b$, is shown in Figure \ref{}.
% We see here that both coefficients are positive and vanish simultaneously at the critical value of the field, $b_c \simeq 0.8$. 
% It is worthwhile to compare  the stiffness coefficient 
% \[ {\cal A} = \sqrt {A_{\|} A_{\perp}}/(2\pi) \]  in Eq.\ \eqref{Esmallq} with the  dispersion law of uniform ferromagnet,  $f_0=0$. The latter case corresponds to $\epsilon_{\mathbf{q}} = q^2 + b $, i.e.\ to ${\cal A}=1$ and the gapped character of the spectrum, which is verified by setting $U=b$, $V=0$ and $\mathbf{A} =0 $ in \eqref{eq:Lagr2}, see Eqs.\  (16)-(18) in \cite{Timofeev_2022}. 

The retarded Green's function is defined as 
 \begin{equation}     \begin{aligned}
     G(t,\mathbf{q})  &= -i\vartheta(t)
     \begin{pmatrix} [u^x_\mathbf{-q} (t),u^x_\mathbf{q}] , & [u^x_\mathbf{-q} (t),u^y_\mathbf{q}] \\ 
[u^y_\mathbf{-q} (t),u^x_\mathbf{q}] , & [u^y_\mathbf{-q} (t),u^y_\mathbf{q}]         
     \end{pmatrix}    \,, 
    \end{aligned}    
\end{equation}
and its Fourier transform at $t>0$ is given by 
\[ %  \begin{equation}     \begin{aligned} 
     G(t,\mathbf{r})     =  -\frac{\sqrt{3} a^2}{4\pi} % -\frac{\vartheta(t)}{2\pi}
     \int \frac{d^2 \mathbf{q}}{(2\pi)^2}
     e^{  i \mathbf{qr} }
     U_1  \begin{pmatrix}  \tfrac{1}{\varkappa}  \sin \epsilon_{\mathbf{q}}t , & -  \cos \epsilon_{\mathbf{q}}t  \\ 
  \cos \epsilon_{\mathbf{q}}t ,  &  \varkappa  \sin \epsilon_{\mathbf{q}}t  \end{pmatrix}  U_1^\dagger 
     \,,    
\]  %  \end{aligned}    \end{equation}
\begin{figure}[t]
\center{\includegraphics[width=0.99\linewidth]{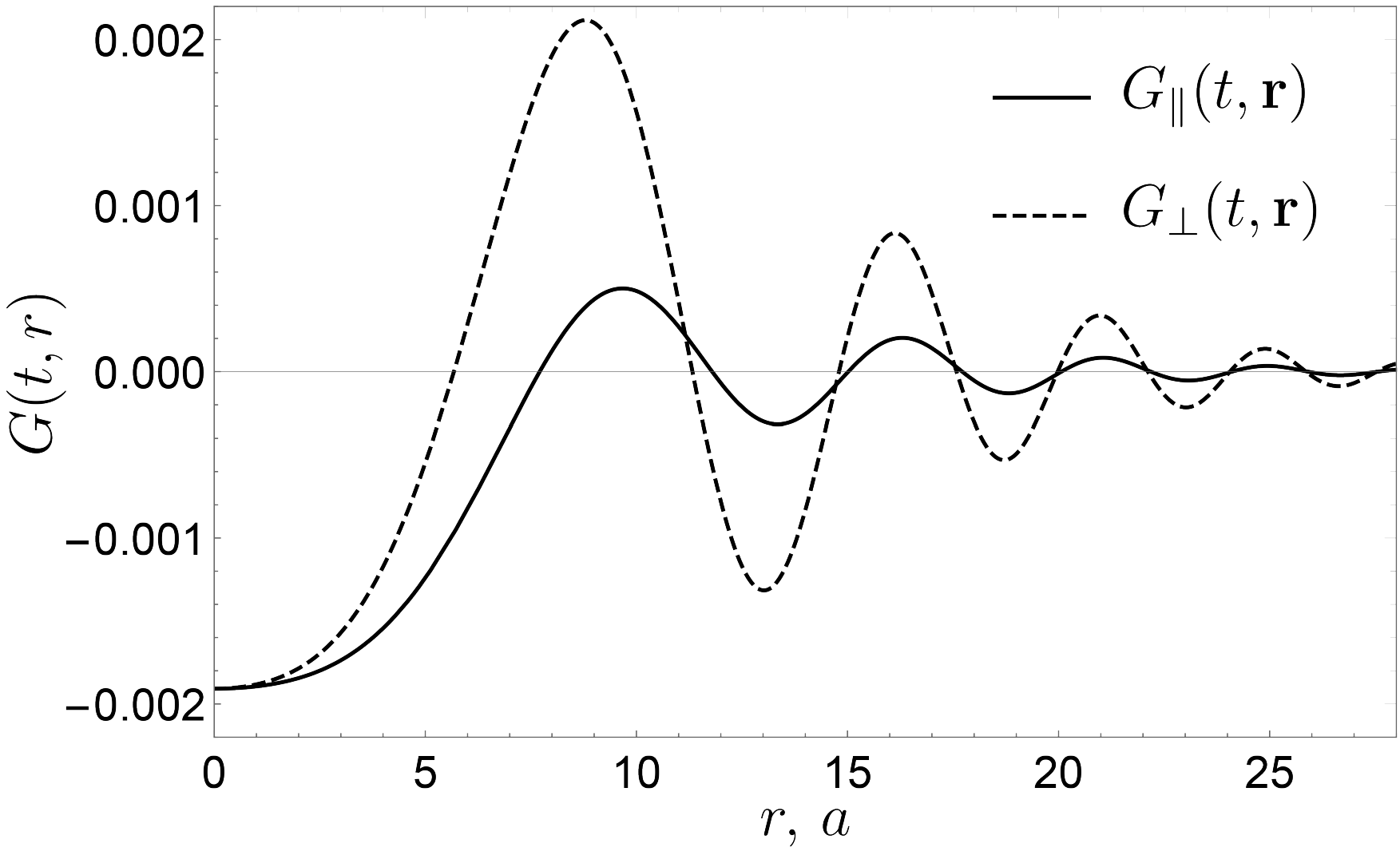}}
\caption{ {Green's functions,  $G_\|(t,\mathbf{r})$ and  $G_\perp(t,\mathbf{r})$, 
plotted for $t=6 {\cal A}^{-1}a^{2}$, 
describing pairwise correlation for displacements, $u^\|$ and $u^\perp$, parallel and perpendicular to $\mathbf{r}$.}
}
\label{fig:corr}
\end{figure}
A simple calculation (effectively restricting the integration over $q\leq a^{-1}$ by the Gaussian form, $\exp{(-q^2 a^2)}$) leads to the expression,  valid for large times and distances, $t{\cal A} \gg r a$, $ r \gg a $, 
\begin{equation}     \begin{aligned} 
     % G(t,\mathbf{r})   &  =   
     % -\frac{\cos (r^2/4 t{\cal A})}{4\pi t{\cal A} } \frac{\kappa + \kappa^{-1}}{2}
     %  \begin{pmatrix}   1  , &0  \\  0 ,  &  1 \end{pmatrix}
     %    \\ 
     %  & + \frac{\sin (r^2/4 t{\cal A})}{4\pi t{\cal A} }
     %  \begin{pmatrix}   0  , & -1  \\  1 ,  &  0 \end{pmatrix}    
     %      \\ 
     % & + \frac{F (r^2/4 t{\cal A})}{4\pi t{\cal A} }\frac{\kappa - \kappa^{-1}}{2}
     %  \begin{pmatrix}   \cos 2\phi  , & \sin 2\phi \\  \sin 2\phi ,  &  -\cos 2\phi \end{pmatrix}  
     %    \,,        \\ 
     %     G(t,\mathbf{r})   &  =  \frac{1}{4\pi t{\cal A}} \left[ 
     % -\cos \left(\frac{r^2}{4 t{\cal A}}\right) \frac{\kappa + \kappa^{-1}}{2}
     %  \begin{pmatrix}   1  , &0  \\  0 ,  &  1 \end{pmatrix} \right.
     %    \\ 
     %  & + \sin \left(\frac{r^2}{4 t{\cal A}}\right)
     %  \begin{pmatrix}   0  , & -1  \\  1 ,  &  0 \end{pmatrix}    
     %      \\ 
     % & \left.  +  F \left(\frac{r^2}{4 t{\cal A}}\right) \frac{\kappa - \kappa^{-1}}{2}
     %  \begin{pmatrix}   \cos 2\phi  , & \sin 2\phi \\  \sin 2\phi ,  &  -\cos 2\phi \end{pmatrix}  
     %  \right] 
     %    \,,        \\ 
         G(t,\mathbf{r})   &  =  \frac{\sqrt{3}a^2}{16\pi^2 t{\cal A}} \left[ 
      - \cos (r^2/4 t{\cal A})\frac{\varkappa + \varkappa^{-1}}{2}
      \begin{pmatrix}   1  , &0  \\  0 ,  &  1 \end{pmatrix} \right.
        \\ 
      & + \sin  (r^2/4 t{\cal A})
      \begin{pmatrix}   0  , & 1  \\  -1 ,  &  0 \end{pmatrix}    
          \\ 
     & \left.  +  F  (r^2/4 t{\cal A}) \frac{\varkappa - \varkappa^{-1}}{2}
      \begin{pmatrix}   \cos 2\phi  , & \sin 2\phi \\  \sin 2\phi ,  &  -\cos 2\phi \end{pmatrix}  
      \right] 
        \,,        \\         
     F (z) &= \cos z - \sin z/z \,, 
    \end{aligned}              
\end{equation}
% In the long time limit, $t\gg r^2/ {\cal A}$, the first term in $G(t,\mathbf{r})$ is dominant, indicating  the isotropic propagation. At intermediate  times, $ra \ll t{\cal A} \ll r^2 $, the combination of the first and third terms shows the anisotropy of tensor $ G(t,\mathbf{r})$, with the main axes along and perpendicular to vector $\mathbf{r}$ in plane.  For shorter times $t{\cal A} \ll r a $, the function $G(t,\mathbf{r})$ becomes (exponentially) small, meaning that the propagation has not yet reached the point $\mathbf{r}$. Notice that the oscillating factors depend only on the ratio $r^2/t$, and the mentioned anisotropy concerns the relative weight of correlations of $u^x$ and $u^y$.

At shorter distances, $ r  \ll \sqrt{t {\cal A}} $, the first term in $G(t,\mathbf{r})$ is dominant, indicating  the isotropic propagation. At intermediate  distances, $  \sqrt{t {\cal A}} \ll r \ll t{\cal A}/a  $, the combination of the first and third terms shows the anisotropy of tensor $ G(t,\mathbf{r})$, with the main axes along and perpendicular to vector $\mathbf{r}$ in plane. For long distances, $r \gg t{\cal A} /a  $, the function $G(t,\mathbf{r})$ becomes (exponentially) small, meaning that the propagation has not yet reached the point $\mathbf{r}$. Notice that the oscillating factors depend only on the ratio $r^2/t$, and the mentioned anisotropy concerns the relative weight of correlations of $u^x$ and $u^y$. We illustrate this behavior  in Fig.\ \ref{fig:corr} by plotting  two principal components,  $G_\|(t,\mathbf{r})$ and  $G_\perp(t,\mathbf{r})$, for pairwise correlations of $u^\|$ and $u^\perp$, respectively. 

{\bf Conclusions}.
We develop a theory of the lowest lying Goldstone mode of the skyrmion lattice, also known as gyrotropic mode. This mode describes the displacements of skyrmions as whole objects and leads to equation of motion in the form of collective Thiele equation. The spectrum is quadratic at small wavevectors, 
and this property 
stems in our approach from the elastic form of the potential and Berry phase kinetic term of the action. The elastic potential follows  from the treatment of skyrmions as individual topological objects and does not assume more demanding theoretical description of phasons, three magnetic helices etc. On the same ground, the quadratic character of the spectrum is robust to inclusion of dipolar interaction or anisotropies as long as SkX is intact. 
The propagation of perturbation through the  skyrmion lattice is anisotropic at intermediate distances. The width of this lowest band monotonically decreases with the magnetic field and disappears at the critical field, marking the transition to the uniform ferromagnetic state.

{\bf Acknowledgements}. 
The work was supported by the Russian Science Foundation, Grant No. 22-22-20034 and St.Petersburg Science Foundation, Grant No. 33/2022. 
The work of  V.T. was partially supported by the Foundation for the Advancement of Theoretical Physics BASIS. % (grant No. 20-1-5-126-1).

\bibliography{skyrmionbib}
\end{document}